\documentclass{ws-ijmpa}

\usepackage[super,compress]{cite}
\usepackage{graphicx}
\usepackage{amsmath,amsfonts,amssymb,dsfont,bm,bbm,dcolumn,float,ifthen,mathrsfs,wasysym}
\usepackage{pstricks,feyn,verbatim,slashed,xcolor,multirow,array,rotating}
\usepackage{url,hyperref,fancybox,epstopdf}

\begin{document}

\markboth{Ning Chen}{Heavy Neutral Higgs Searches}

%
\catchline{}{}{}{}{}
%


\newcommand{\keV}   {~\mathrm{keV}}
\newcommand{\GeV}      {~\mathrm{GeV}}
\newcommand{\TeV}      {~\mathrm{TeV}}
\newcommand{\MeV}      {~\mathrm{MeV}}

\newcommand{\pb}      {~\mathrm{pb}}
\newcommand{\fb}      {~\mathrm{fb}}
\newcommand{\ab}      {~\mathrm{ab}}

\newcommand{\Tr}   {~\mathrm{Tr}}

\newcommand{\ba}{\begin{array}}
\newcommand{\ea}{\end{array}}
\newcommand{\beqn}{\begin{eqnarray}}
\newcommand{\eeqn}{\end{eqnarray}}
\newcommand{\beqs}{\begin{subequations}}
\newcommand{\eeqs}{\end{subequations}}
\newcommand{\be}{\begin{equation}}
\newcommand{\ee}{\end{equation}}
\newcommand{\non}{\nonumber \\}

\def\gU{\rm U}
\def\gSU{\rm SU}
\def\gO{\rm O}
\def\gSO{\rm SO}
\def\gSp{\rm Sp}
\def\gUSp{\rm USp}
\def\gE{\rm E}
\def\gF{\rm F}
\def\gG{\rm G}

\def\mA{\mathcal{A}}
\def\mB{\mathcal{B}}
\def\mC{\mathcal{C}}
\def\mD{\mathcal{D}}
\def\mE{\mathcal{E}}
\def\mF{\mathcal{F}}
\def\mG{\mathcal{G}}
\def\mH{\mathcal{H}}
\def\mI{\mathcal{I}}
\def\mJ{\mathcal{J}}
\def\mK{\mathcal{K}}
\def\mL{\mathcal{L}}
\def\mM{\mathcal{M}}
\def\mN{\mathcal{N}}
\def\mO{\mathcal{O}}
\def\mP{\mathcal{P}}
\def\mQ{\mathcal{Q}}
\def\mR{\mathcal{R}}
\def\mS{\mathcal{S}}
\def\mT{\mathcal{T}}
\def\mU{\mathcal{U}}
\def\mV{\mathcal{V}}
\def\mW{\mathcal{W}}
\def\mX{\mathcal{X}}
\def\mY{\mathcal{Y}}
\def\mZ{\mathcal{Z}}

\def\hf{\frac{1}{2}}

\newcommand{\vecMET}{\vec E\hspace{-0.08in}\slash_T}
\newcommand{\MET}{E\hspace{-0.08in}\slash_T}

\title{The LHC Searches for Heavy Neutral Higgs Bosons by Jet Substructure Analysis}

\author{Ning Chen}
\address{Department of Modern Physics, University of Science and Technology of China\\
Hefei, Anhui 230026, China\\
chenning@ustc.edu.cn}


\maketitle


\begin{abstract}

The two-Higgs-doublet model contains extra Higgs bosons, with mass ranges spanning from several hundred GeV to about 1 TeV. 
We study the possible experimental searches for the neutral Higgs bosons of $A$ and $H$ at the future high-luminosity LHC runs. 
Besides of the conventional search modes that are inspired by the supersymmetric models, we discuss two search modes which were not quite addressed previously. 
They are the decay modes of $A\to hZ$ and $A/H \to t \bar t$. 
Thanks to the technique of tagging boosted objects of SM-like Higgs bosons and top quarks, we show the improved mass reaches for heavy neutral Higgs bosons with masses up to $\sim\mO(1)\,\TeV$.
The modes proposed here are complementary to the conventional experimental searches motivated by the MSSM.

\end{abstract}

\ccode{PACS numbers:12.60.Fr, 14.80.-j, 14.80.Ec}



\section{Introduction}
\label{section:introduction}

In many of new physics models beyond the SM (BSM), the Higgs sector is extended with several scalar multiplets. 
Examples include the minimal supersymmetric standard model (MSSM)~\cite{Dimopoulos:1981zb}\,, the left-right symmetric models~\cite{Zhang:2007da}\,, and the composite Higgs models~\cite{Mrazek:2011iu}\,. 
There are several Higgs bosons in these models, with one of them to be identified as the $125\,\GeV$ Higgs boson discovered at LHC. 
Therefore, extra heavy Higgs bosons are yet to be searched for by the future LHC experiments and the future high-energy $pp$ colliders running at $\sqrt{s}=50 - 100\,\TeV$~\cite{Gomez-Ceballos:2013zzn, CEPC-SPPC-pre}\,.

A very widely studied scenario beyond the minimal one-doublet setup is the two-Higgs-doublet model (2HDM), which is the low-energy description of the scalar sectors in various new physics models. 
A recent review of the phenomenology in the context of the general 2HDM can be found in Ref.~\citen{Branco:2011iw}. Refs.~\citen{Craig:2012vn, Craig:2012pu, Coleppa:2013dya, Craig:2013hca, Coleppa:2013xfa,Carena:2013ooa, Chen:2013emb, Chen:2014dma, Baglio:2014nea, Coleppa:2014hxa, Dumont:2014wha, Dorsch:2014qja, Hespel:2014sla, Barger:2014qva, Fontes:2014xva, Coleppa:2014cca, Grzadkowski:2014ada, Gunion:2012he} studied the 2HDM phenomenology at the LHC in light of the $125\,\GeV$ Higgs discovery. 
The scalar spectrum in the 2HDM contains five states, namely, two neutral $CP$-even Higgs bosons $(h\,,H)$, one neutral $CP$-odd Higgs boson $A$, and two charged Higgs bosons $H^\pm$. 
In the context of the general 2HDM, each Higgs boson mass is actually a free parameter before applying any constraint. 
By including the perturbative unitarity and stability constraints to the general 2HDM potential~\cite{Barger:2014qva}\,, the masses of the heavy Higgs bosons in the spectra are generally bounded from above as $(M_A\,,M_H\,,M_\pm)\lesssim 1\,\TeV$.
Therefore, it becomes evident that the upcoming LHC runs at 14 TeV would search for these heavy states in the mass range of several hundred GeV to $\sim\mO(1)\,\TeV$.

Within the framework of the 2HDM, we study the high-luminosity (HL) LHC searches for the heavy neutral Higgs bosons $A$ and $H$ at $14\,\TeV$ run. 
The previous experimental searches often focus on the benchmark models in the MSSM, which has type-II 2HDM Yukawa couplings. 
Thus, the interesting final states to be looked for are the $A/H\to \bar b b$~\cite{Aaltonen:2012zh,Chatrchyan:2013qga} and $A/H\to \tau^+ \tau^-$~\cite{Schael:2006cr,Abazov:2008hu,Aaltonen:2009vf, Abazov:2011up,Chatrchyan:2012vp, Aad:2012cfr, Aad:2014vgg} due to the significant enhancements to the Yukawa couplings.
Different from the existing experimental search modes, we consider the decay modes of $A/H\to t \bar t$ and $A \to hZ$.
The $A/H\to t \bar t$ decay mode can be the most dominant one with the low-$t_\beta$ inputs, for both 2HDM-I and 2HDM-II setups. 
Due to the large SM background of $t \bar t$, the searches for the $t \bar t$ final states from the heavy Higgs boson decays are thought to be very challenging. 
In addition, it is known that the signal channel of $gg\to A/H \to t \bar t $ strongly interferes with the SM background~\cite{Dicus:1994bm, Frederix:2007gi, Jung:2015gta} and results in a peak-dip structure. 
Therefore, one can only rely on the heavy-quark associated production channels to study the $A/H \to t \bar t $ decays. 
For the $A\to hZ$ decay modes, we study the $\bar b b\ell^+ \ell^-$ final state searches, where a SM-like Higgs boson with mass of $125\,\GeV$ is involved.

A common feature is that both decay modes involve heavy states.
With heavy mother particles of $A/H$, it is natural to expect large boosts to the SM-like Higgs boson $h$ and/or top quarks in our study.
For highly boosted top quarks and/or $h$, the jets in their hadronic decay modes may lie close together and may not be independently resolved.
As a result, the top quarks and/or SM-like Higgs boson $h$ in the boosted region may appear as single jets with three or two subjets in a small region of the calorimeter.
The separations by angular scales of subjets are of order $2m_h/p_T$ or $2m_t/p_T$.
The method of tagging the boosted SM-like Higgs jets was suggested in Refs.~\citen{Butterworth:2008iy, Butterworth:2008tr}, where one is likely to use the dominant decay mode of $h_{\rm SM}\to b \bar b$.
This is dubbed ``BDRS'' algorithm.
The method of reconstructing the boosted top jet uses the hadronic decay mode of $t_h\to b W_h \to b + jj$.
Specifically, there are two classes of methods of tagging the boosted top quarks.
One algorithm is called the {\sc JHUTopTagger}~\cite{Kaplan:2008ie}\,, which requires the summation of the transverse momenta of the decayed particles to be larger than $1\,\TeV$. 
This is very challenging when one is interested in mother particles with masses of several hundred $\GeV$ to $\mO(1)\,\TeV$. 
Alternatively, we study the LHC searches for the $A/H \to t \bar t$ decays by using the {\sc HEPTopTagger} method~\cite{Plehn:2009rk, Plehn:2010st, Plehn:2011sj, Plehn:2011tg}, which is efficient in tagging the top jets with intermediate transverse momenta of $\mO(100)\,\GeV$.

The rest of the article is organized as follows. 
In Sec.~\ref{section:AHin2HDM}, we have a brief review of the heavy neutral Higgs bosons in the framework of the general $CP$-conserving 2HDM. 
In the precise alignment limit of $c_{\beta - \alpha }\to 0$, the decay modes of $A/H\to t  \bar t$ are always the most dominant ones for the 2HDM-I, and also dominant ones for the 2HDM-II with low-$t_\beta$ inputs.  
The inclusive production cross sections of $\sigma[pp\to t \bar t + (A/H \to t \bar t) ]$ at the LHC $14\,\TeV$ runs are evaluated. 
The relaxation of the alignment limit leads to possible exotic decay modes, such as $A\to hZ$.
We show that this decay mode can also become significant, especially for the 2HDM-I case.
In Sec.~\ref{section:AHtt}, we search for the $t \bar t + ( A/H\to t \bar t) $ channel~\cite{Chen:2015fca} at the HL-LHC. 
We focus on the $t\bar t + A/H$ production channel, with the sequential decay of $A/H\to t \bar t$.
This process is always controlled by the top quark Yukawa coupling of the heavy Higgs bosons. 
By applying the {\sc HEPTopTagger} method for reconstructing one boosted top quark, plus selecting two additional same-sign-dilepton (SSDL) events, we obtain a signal reach for $M_{A/H}\sim \mO(1)\,\TeV$ with low-$t_\beta$ inputs at the HL-LHC runs.
The results in this part are mainly based on the Ref.~\citen{Chen:2015fca}.
In Sec.~\ref{section:AhZ}, the analysis of LHC searches for the $CP$-odd Higgs boson via the $A\to hZ$ final states~\cite{Chen:2014dma} is provided.
In order to eliminate the SM background sufficiently, we apply the BDRS algorithm in Ref.~\citen{Butterworth:2008iy} to reconstruct the boosted Higgs.
The LHC search potential to the $A\to hZ$ decay channel at different phases of the upcoming runs at $14\,\TeV$ is shown.
The results in this part are mainly based on the Ref.~\citen{Chen:2014dma}.
Finally, we make conclusion in Sec.~\ref{section:conclusion}.


\section{The Heavy Neutral Higgs Bosons in The 2HDM}
\label{section:AHin2HDM}

In this section, we briefly discuss the productions and decays of the heavy neutral Higgs bosons $A$ and $H$ in the context of the general $CP$-conserving 2HDM.


\subsection{The general 2HDM setup and couplings}

The most general 2HDM Higgs potential is composed of all gauge-invariant and renormalizable terms by two Higgs doublets $(\Phi_1\,, \Phi_2)\in 2_{+1}$ of the $\gSU(2)_L \times \gU(1)_Y$ electroweak gauge symmetries. 
For the $CP$-conserving case, there can be two mass terms plus seven quartic coupling terms with real parameters. 
For simplicity, we consider the soft breaking of a discrete $\mathbb{Z}_{2}$ symmetry, under which two Higgs doublets transform as $(\Phi_1\,,\Phi_2)\to (-\Phi_1\,, \Phi_2)$. 
The corresponding Lagrangian is expressed as
\beqn\label{eq:2HDM_potential}
\mL&=&\sum_{i=1\,,2}|D \Phi_i|^2 - V(\Phi_1\,,\Phi_2)\,,\\
V(\Phi_1\,,\Phi_2)&=&m_{11}^2|\Phi_1|^2+m_{22}^2|\Phi_2|^2-m_{12}^2(\Phi_1^\dag\Phi_2+H.c.)+\hf\lambda_1 |\Phi_{1}|^{4} +\hf\lambda_2|\Phi_{2}|^{4}\non
&+&\lambda_3|\Phi_1|^2 |\Phi_2|^2+\lambda_4 |\Phi_1^\dag \Phi_2|^2+\hf\lambda_5 \Big[ (\Phi_1^\dag\Phi_2) (\Phi_1^\dag\Phi_2)+H.c.\Big]\,.
\eeqn
Two Higgs doublets $\Phi_1$ and $\Phi_2$ pick up VEVs to trigger the EWSB,
%
%
and one parametrizes the ratio of the two Higgs VEVs as
\beqn\label{eq:tb}
t_\beta&\equiv&\tan\beta = \frac{v_2}{v_1}\,.
\eeqn
The perturbative bounds of the heavy Higgs boson Yukawa couplings constrain the choices of $t_\beta$, which should be neither as small as $\mO(0.1)$ nor as large as $\mO(50)$. 
Three of the eight real components correspond to the Nambu-Goldstone bosons giving rise to the electroweak gauge boson masses, with the remaining five as the physical Higgs bosons, namely, two $CP$-even Higgs bosons $h$ and $H$, one $CP$-odd Higgs boson $A$, and the charged Higgs bosons $H^\pm$. 
The light $CP$-even Higgs boson $h$ is taken as the only state in the 2HDM spectra with mass of $125\,\GeV$ and its couplings with SM fermions and gauge bosons are controlled by two parameters of $(\alpha\,,\beta)$. 
A more convenient choice of 2HDM parameter set is $(c_{\beta-\alpha}\,, t_\beta)$. 
The current global fits~\cite{Coleppa:2013dya, Craig:2013hca, Barger:2013ofa,Chowdhury:2015yja} by using the LHC $7\oplus 8\,\TeV$ runs to the 2HDM parameters point to the alignment limit of $c_{\beta- \alpha}\to 0$.

 \begin{table}[ph]
\tbl{The Yukawa couplings of the heavy neutral Higgs bosons $H$ and $A$ in the general 2HDM.}
{\begin{tabular}{@{        }ccc@{        }} \toprule
 & 2HDM-I &  2HDM-II \\
 \colrule
  $\xi_H^{u}$  & $c_{\beta-\alpha} -s_{\beta-\alpha}/t_{\beta}$ & $c_{\beta-\alpha} -s_{\beta-\alpha}/t_{\beta}$  \\
 $\xi_H^{d}$  & $c_{\beta-\alpha} -s_{\beta-\alpha}/t_{\beta}$ & $c_{\beta-\alpha} + t_{\beta} s_{\beta-\alpha}$  \\
 $\xi_H^{\ell}$  & $c_{\beta-\alpha} -s_{\beta-\alpha}/t_{\beta}$ & $c_{\beta-\alpha} + t_{\beta} s_{\beta-\alpha} $ \\ \colrule
 $\xi_{A}^{u}$  & $1/t_{\beta}$ & $1/t_{\beta}$  \\
 $\xi_{A}^{d}$  & $-1/t_{\beta}$ & $t_{\beta}$  \\
 $\xi_{A}^{\ell}$  & $-1/t_{\beta}$ & $t_{\beta}$ \\  \botrule
\end{tabular} \label{table:AH_yuk}}
\end{table}

In the general 2HDM, SM fermions with the same quantum numbers couple to the same Higgs doublet, which will avoid the tree-level flavor-changing neutral currents. 
For the 2HDM-I, all SM fermions couple to one Higgs doublet (conventionally chosen to be $\Phi_2$). This setup can be achieved by assigning a discrete $\mathbb{Z}_2$ symmetry under which $\Phi_1 \to - \Phi_1$. 
For the 2HDM-II, the up-type quarks $u_i$ couple to one Higgs doublet (conventionally chosen to be $\Phi_2$) and the down-type quarks $d_i$ and the charged leptons $\ell_i$ couple to the other ($\Phi_1$). 
This can also be achieved by assigning a discrete $\mathbb{Z}_2$ symmetry under which $\Phi_1 \to - \Phi_1$ together with $(d_i\,, \ell_i)\to (- d_i \,, - \ell_i)$. 
At the tree level, the Yukawa coupling terms for neutral Higgs bosons are expressed as
\beqn\label{eq:Yukawa}
-\mL_Y&=&\sum_f  \frac{m_f}{v} ( \xi_H^f  \bar f f H -i\xi_A^f \bar f \gamma_5 f A)\,.
\eeqn
%
The dimensionless coupling strengths of $\xi_H^f$ and $\xi_A^f$ are listed in Table.~\ref{table:AH_yuk} for the 2HDM-I and 2HDM-II cases, respectively. 
In the alignment limit, all dimensionless Yukawa couplings of $\xi_H^f$ and $\xi_A^f$ are inversely proportional to $t_\beta$ for the 2HDM-I, while they depend on the $t_\beta$ inputs in the manner of $\xi_{A/H}^u\propto 1/t_\beta$ and $\xi_{A/H}^d \propto t_\beta$ for the 2HDM-II.
With the low-$t_\beta$ inputs, the heavy neutral Higgs bosons $A$ and $H$ always couple strongly to the top quarks.

Besides of Yukawa couplings, there are also other couplings relevant to the decays of the heavy neutral Higgs bosons $A$ and $H$. 
From the 2HDM kinematic terms $|D \Phi_i |^2$, one has Higgs-gauge couplings of $G(HVV)$, $G(AhZ/AHZ)$ and $G(A H^\pm W^\mp)$. 
The $G(HVV)$ and $G(AhZ)$ couplings are vanishing in the $c_{\beta - \alpha}=0$ limit. 
From the general 2HDM potential, one also has the triple Higgs couplings such as $G(Hhh)$, $G(HAA)$, and $G(H H^+ H^-)$. 
The existences of these couplings lead to exotic heavy Higgs boson search strategies. 
Though we focus on the alignment limit of $c_{\beta-\alpha}=0$, it is noted that the current global fits to the $125\,\GeV$ $CP$-even Higgs boson $h$ in the 2HDM generally allow the parameter choices of $c_{\beta-\alpha}\sim \mO(0.1)$ for 2HDM-I and $c_{\beta-\alpha}\sim \mO(0.01)$ for 2HDM-II, respectively. 
It was shown in the previous discussions~\cite{Chen:2013emb, Chen:2014dma} that some of the heavy Higgs boson decay modes of $A\to hZ$ and $H\to hh$ can become the leading ones, especially for $M_{A/H}\lesssim 2 m_t$. 
The relevant LHC searches can be performed by the boosted Higgs searches plus the opposite-sign-same-flavor (OSSF) di-leptons, and via the $b \bar b+ \gamma\gamma$ final states.


\subsection{The productions of $A$ and $H$}

Since the global fits to the general 2HDM point to the alignment limit of $c_{\beta -\alpha }=0$, the production channels of the heavy neutral Higgs bosons $A/H$ at the LHC are most likely due to the gluon fusion and the heavy quark associated processes~\cite{Djouadi:2005gi, Djouadi:2005gj}\,.
The VBF and vector boson associated processes are highly suppressed for the heavy $CP$-even Higgs boson $H$, and they are absent for the $CP$-odd Higgs boson $A$.

\begin{figure}
\centering
\includegraphics[width=5.5cm,height=4cm]{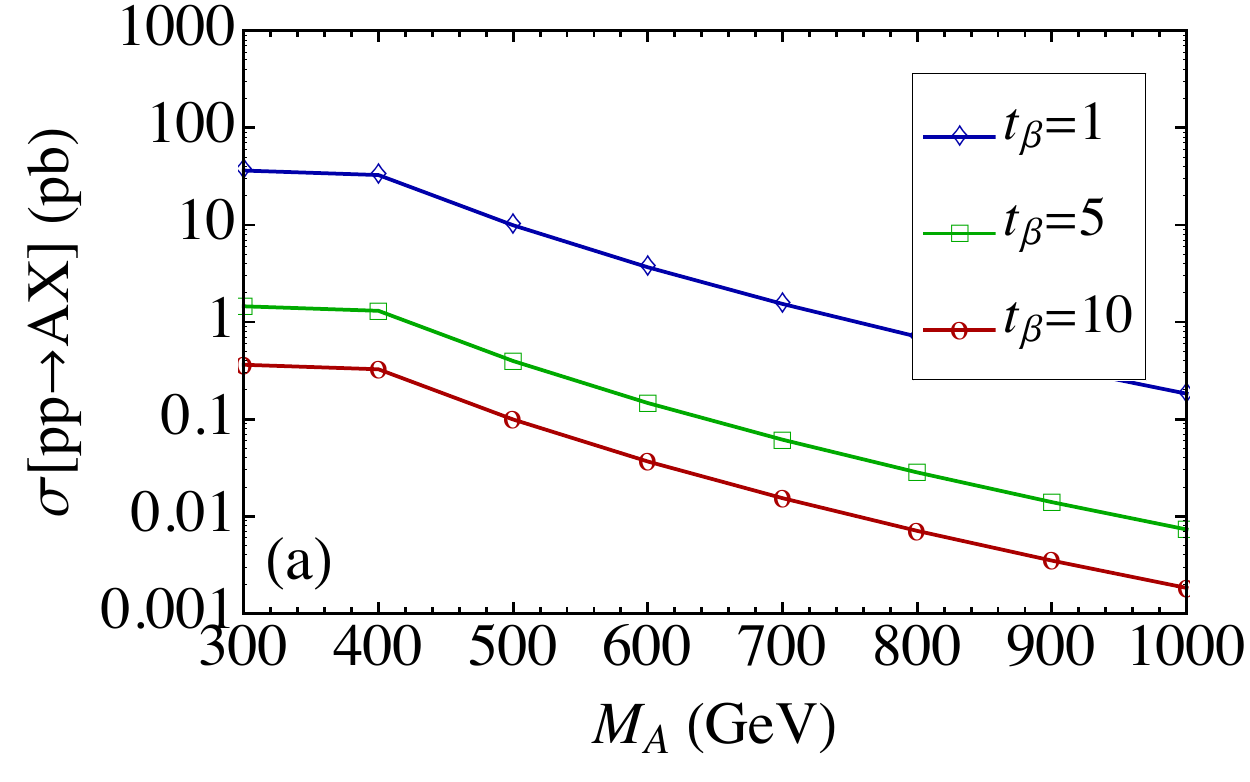}
\includegraphics[width=5.5cm,height=4cm]{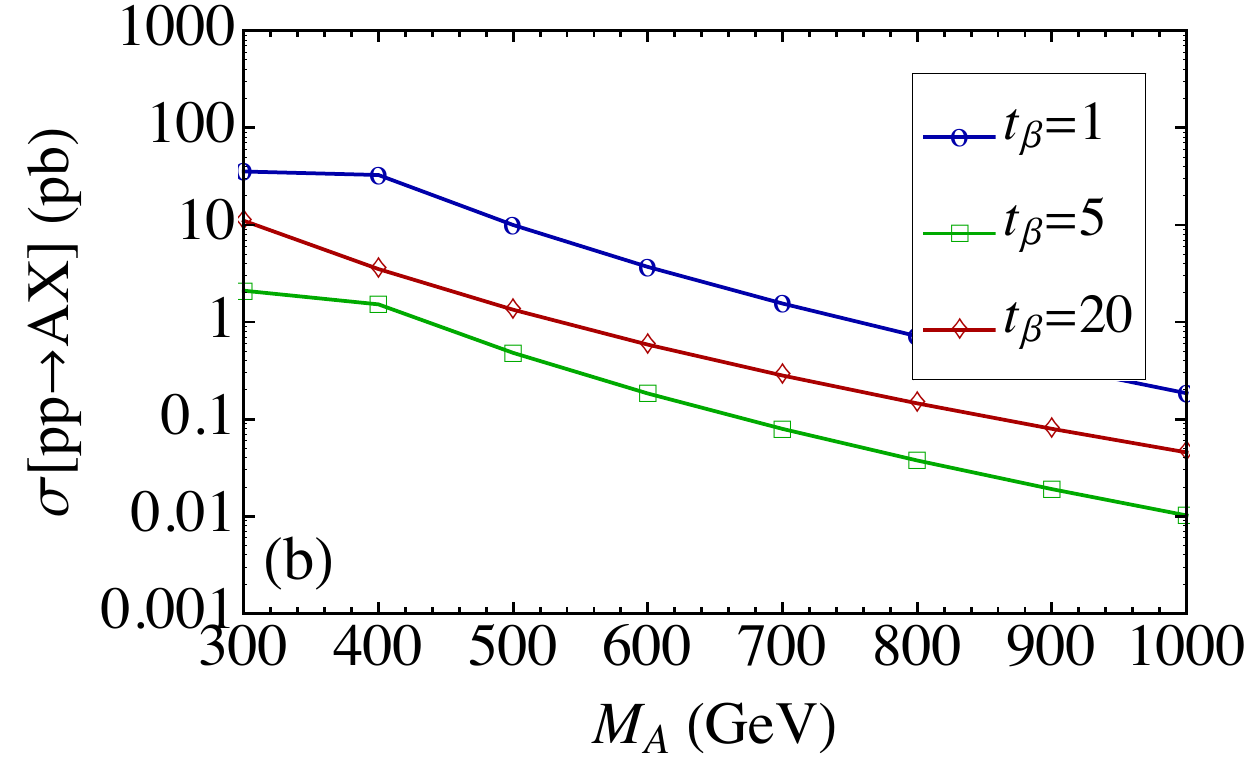}
\caption{\label{fig:pptoA} The inclusive production cross section $\sigma[pp\to AX]$ for $M_A \in (300\,\GeV , 1\,\TeV)$ at the LHC $14\,\TeV$ runs. Left: 2HDM-I, with inputs of $t_\beta=1$ (blue), $t_\beta=5$ (green), and $t_\beta=10$ (red). Right: 2HDM-II, with inputs of $t_\beta=1$ (blue), $t_\beta=5$ (green), and $t_\beta=20$ (red).}
\end{figure}

At the leading order, the parton-level production cross section of $\hat \sigma(gg\to A)$ is related to the gluonic partial decay width as follows:
\beqs
\beqn
\hat \sigma(gg\to A)&=&\frac{\pi^2}{8 M_A} \Gamma[A\to gg] \delta(\hat s - M_A^2)\,,\label{eq:ggA}\\
 \Gamma[A\to gg]&=&\frac{G_F \alpha_s^2 M_A^3}{64\sqrt{2} \pi^3 } \Big| \sum_q \xi_A^q A_{1/2}^A(\tau_q)   \Big|^2\,,\label{eq:Atogg}
\eeqn
\eeqs
with $\tau_{q}\equiv M_A^{2}/(4 m_{q}^{2})$ and $\xi_A^q$ being the Yukawa couplings given in Table.~\ref{table:AH_yuk}. 
Here, $A_{1/2}^A(\tau)$ is the fermionic loop factor for the pseudoscalar. 
In the heavy quark mass limit of $m_q\gg M_A$, this loop factor reaches the asymptotic value of $ A_{1/2}^A(\tau)\to 2$, while it vanishes in the chiral limit of $m_q\ll M_A$. 
In practice, we evaluate the production cross sections for these processes by {\sc SusHi}~\cite{Harlander:2012pb}\,. 
The inclusive production cross sections of $\sigma[pp\to AX]$ are shown in Fig.~\ref{fig:pptoA} for the LHC runs at $14\,\TeV$, where the $CP$-odd Higgs boson is considered in the mass range of $M_A\in (300\,\GeV, 1\,\TeV)$.

For the decay modes of $A/H\to t \bar t$, the previous studies of the gluon fusion to $t \bar t$ final states via the spin-0 resonances have suggested strong interference effects with the QCD backgrounds. 
Therefore, we are left with the heavy-quark associated productions as the possible channels to search for at the LHC.
For the 2HDM-I case, all dimensionless Yukawa couplings of the SM fermions to the $A/H$ are universally proportional to $1/t_\beta$. 
For the 2HDM-II case, the dimensionless Yukawa couplings scale as $\xi_{A/H}^u\propto 1/t_\beta$ and $\xi_{A/H}^d \propto t_\beta$, respectively. 
Therefore, one expects the production cross sections of $\sigma [pp \to  t \bar t + A/H]\sim 1/t_\beta^2 $ to become significant with the low-$t_\beta$ inputs.


\subsection{The decays of $A$ and $H$}

 \begin{table}[ph]
 \tbl{The classification of the $CP$-odd Higgs boson $A$ decay modes in the general 2HDM. A checkmark (dash) indicates that the decay mode is present (absent) in the $c_{\beta-\alpha}=0$ alignment limit. }
{\begin{tabular}{@{        }   cccc   @{        }} \toprule    
 $A$ decays & final states &   & alignment limit \\
 \hline
 \multirow{2}{*}{SM fermions}  & $A\to (\tau^+ \tau^- \,, \mu^+ \mu^-) $  & & $\checkmark$  \\
 & $A\to (t \bar t\,, b \bar b ) $  & & $\checkmark$  \\
\colrule
 \multirow{3}{*}{Exotics}  & $A\to h Z$ &   & $-$ \\
 & $A\to H Z$ &  & $\checkmark$  \\
 & $A\to H^\pm W^\mp$ &   & $\checkmark$ \\
\colrule
 Loops & $A\to( gg\,, \gamma \gamma \,, \gamma Z)$  & & $\checkmark$ \\
\botrule
\end{tabular}\label{tab:Adecay}}
\end{table}

\begin{table}[ph]
\tbl{
The classification of the $CP$-even Higgs boson $H$ decay modes in the general 2HDM. 
A checkmark (dash) indicates that the decay mode is present (absent) in the $c_{\beta-\alpha}=0$ alignment limit. }
{\begin{tabular}{@{        }   cccc   @{        }} \toprule    
 $H$ decays & final states &   & alignment limit \\
\colrule
 \multirow{2}{*}{SM fermions} & $H\to (\tau^+ \tau^- \,, \mu^+ \mu^-) $  & & $\checkmark$  \\
  & $H\to (t \bar t\,, b \bar b) $  & & $\checkmark$  \\
 \colrule
gauge bosons & $H\to (WW\,, ZZ) $  & & $-$  \\
 \colrule
\multirow{5}{*}{Exotics} & $H\to AZ$ &   & $\checkmark$ \\
 & $H\to H^\pm W^\mp$ &   & $\checkmark$ \\
 & $H\to h h$ &   & $-$ \\
 & $H\to AA$ &   & $-$ \\
 & $H\to H^+ H^-$ &  & $\checkmark$ \\
 \colrule
Loops & $H\to(gg\,, \gamma \gamma \,, \gamma Z)$  & & $\checkmark$ \\
\botrule
\end{tabular}\label{tab:Hdecay}}
\end{table}

\begin{figure}
\centering
\includegraphics[width=6.5cm,height=4.5cm]{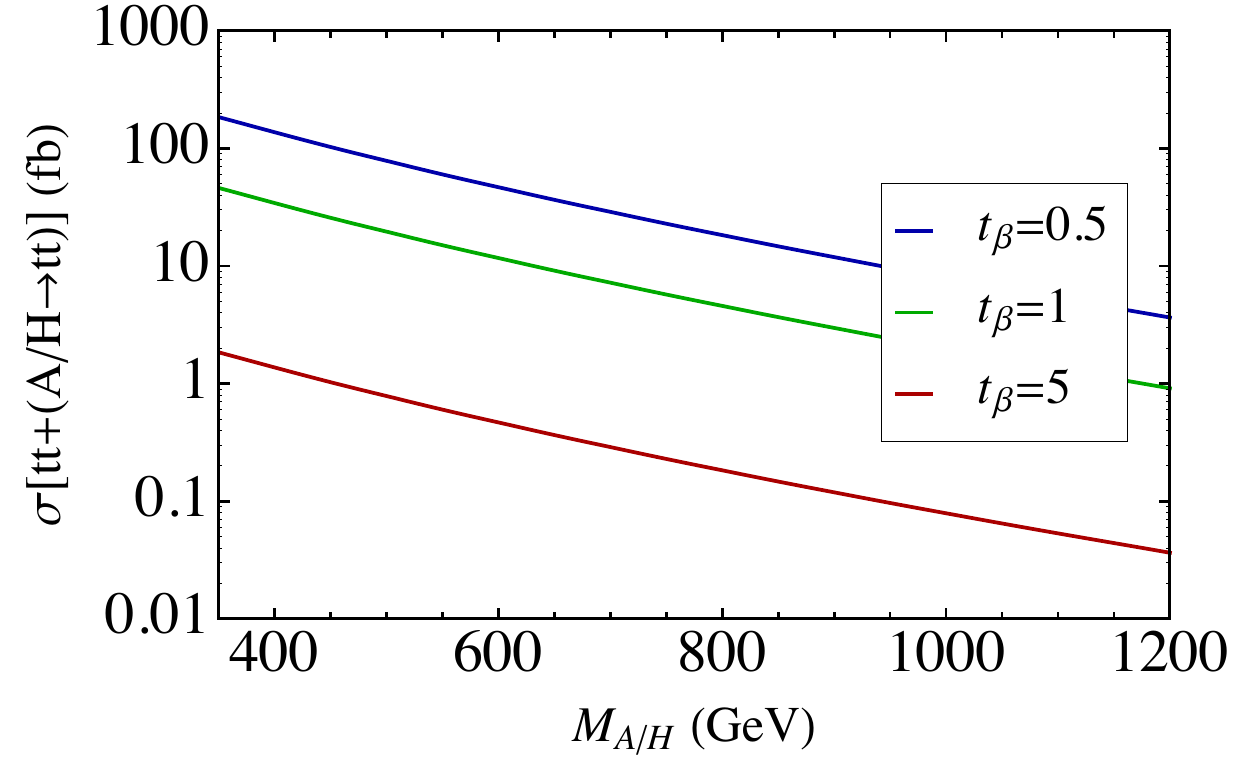}
\caption{\label{fig:AHttxsec} 
$\sigma[pp\to t \bar t A/H]\times {\rm BR}[A/H\to t \bar  t]$ (for both 2HDM-I and 2HDM-II cases) with $M_{A/H}\in (350\,\GeV, 1200\,\GeV)$ at the LHC $14\,\TeV$ runs.}
\end{figure}

All possible decay modes of heavy neutral Higgs boson are listed in Tables.~\ref{tab:Adecay} and \ref{tab:Hdecay} for $A$ and $H$, respectively. 
The presence or absence of these decay modes in the alignment limit are also marked.
In practice, we evaluate their partial decay widths by using {\sc 2HDMC}~\cite{Eriksson:2009ws}\,. 
Some exotic decay modes involving another heavy states, such as $A\to HZ$ and $H\to H^\pm W^\mp$, are always turned off for later discussions.
This is reasonable when one assumes the masses of all heavy Higgs bosons are close to each other.
The loop-induced decay branching ratios of ${\rm Br}[A/H \to \gamma\gamma/ \gamma Z]$ are typically smaller than $10^{-5}$, which can be neglected. 
For the 2HDM-I case, the decay branching ratios of ${\rm Br}[A/H\to t \bar  t]$ are always dominant to be $\sim \mO(1)$ since all dimensionless Higgs Yukawa couplings scale as $\sim 1/t_\beta$. 
For the 2HDM-II case, the ${\rm Br}[A/H\to t \bar  t]$ can be suppressed to $\sim\mO(0.1)$ with the large-$t_\beta$ inputs, where the partial decay widths of $\Gamma [A/H \to b \bar  b]$ and $\Gamma[A/H \to \tau^+ \tau^-]$ become dominant. 
Combining the production cross sections evaluated by {\sc Madgraph 5}~\cite{Alwall:2014hca}\,, we demonstrated the cross sections of $\sigma[pp\to t \bar t+A/H] \times {\rm Br}[A/H\to t \bar t  ]$ within the mass range of $M_{A/H}\in (350\,\GeV\,, 1200\,\GeV)$ at the LHC $14\,\TeV$ runs in Fig.~\ref{fig:AHttxsec}.  
As it turns out, the decay branching ratios of ${\rm Br}[A/H \to t \bar t]$ tend to unity for both 2HDM-I and 2HDM-II with the small-$t_\beta$ inputs of $\sim \mO(1)$. 
For this reason, we combine the cross sections of $\sigma[pp\to t \bar t + A/H] \times {\rm Br}[A/H\to t \bar t  ]$ for both 2HDM-I and 2HDM-II into one plot.

\begin{figure}
\centering
\includegraphics[width=5cm,height=4cm]{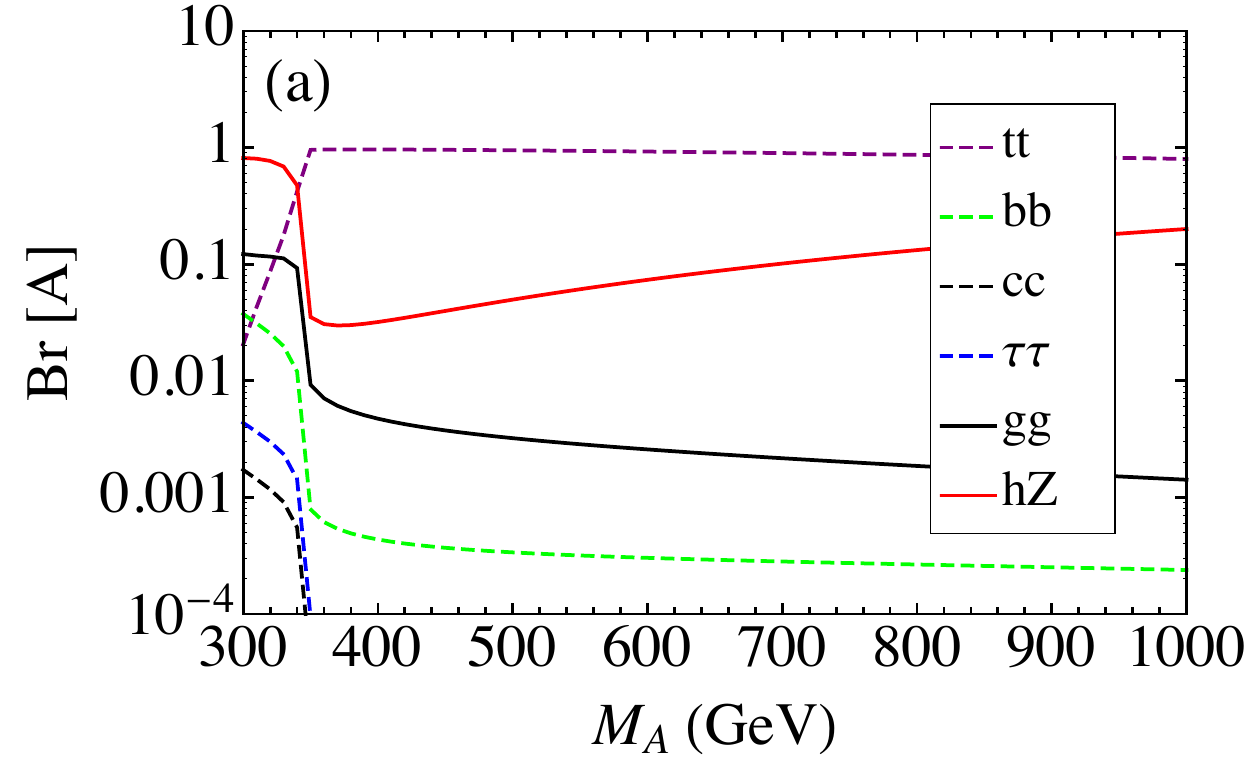}
\includegraphics[width=5cm,height=4cm]{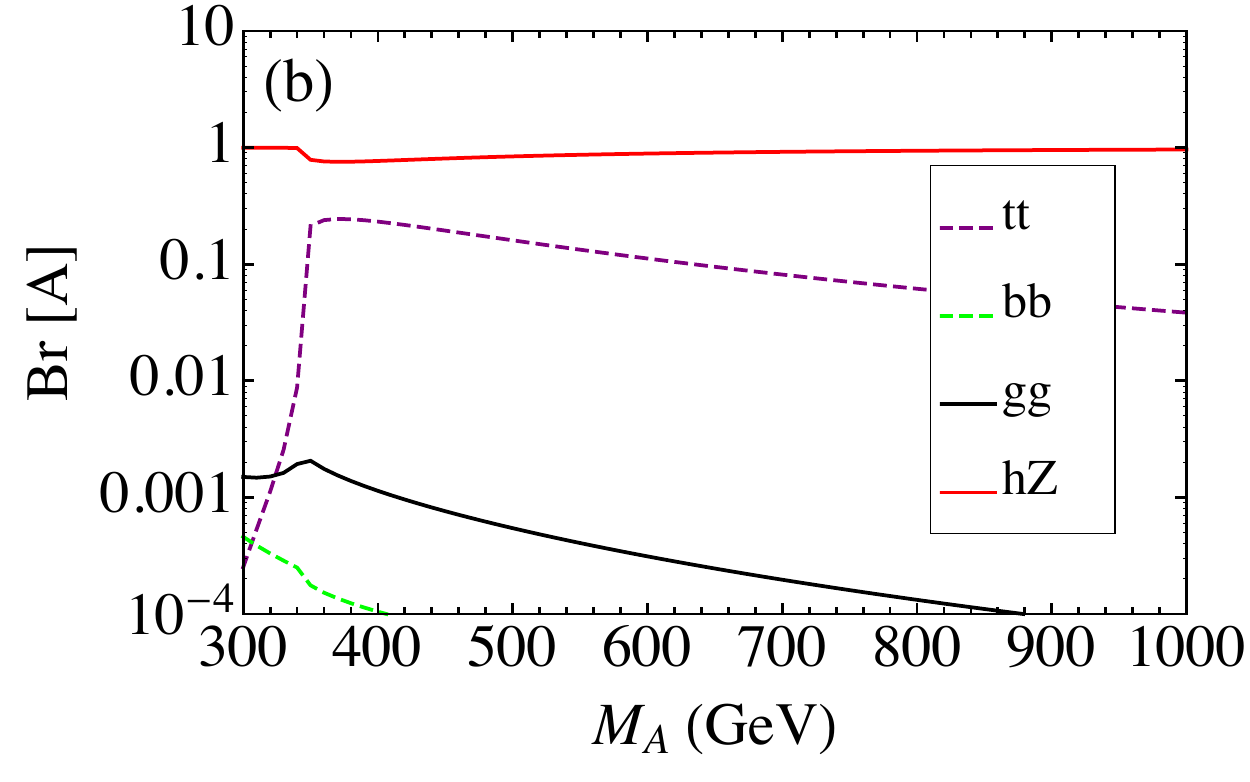}\\
\includegraphics[width=5cm,height=4cm]{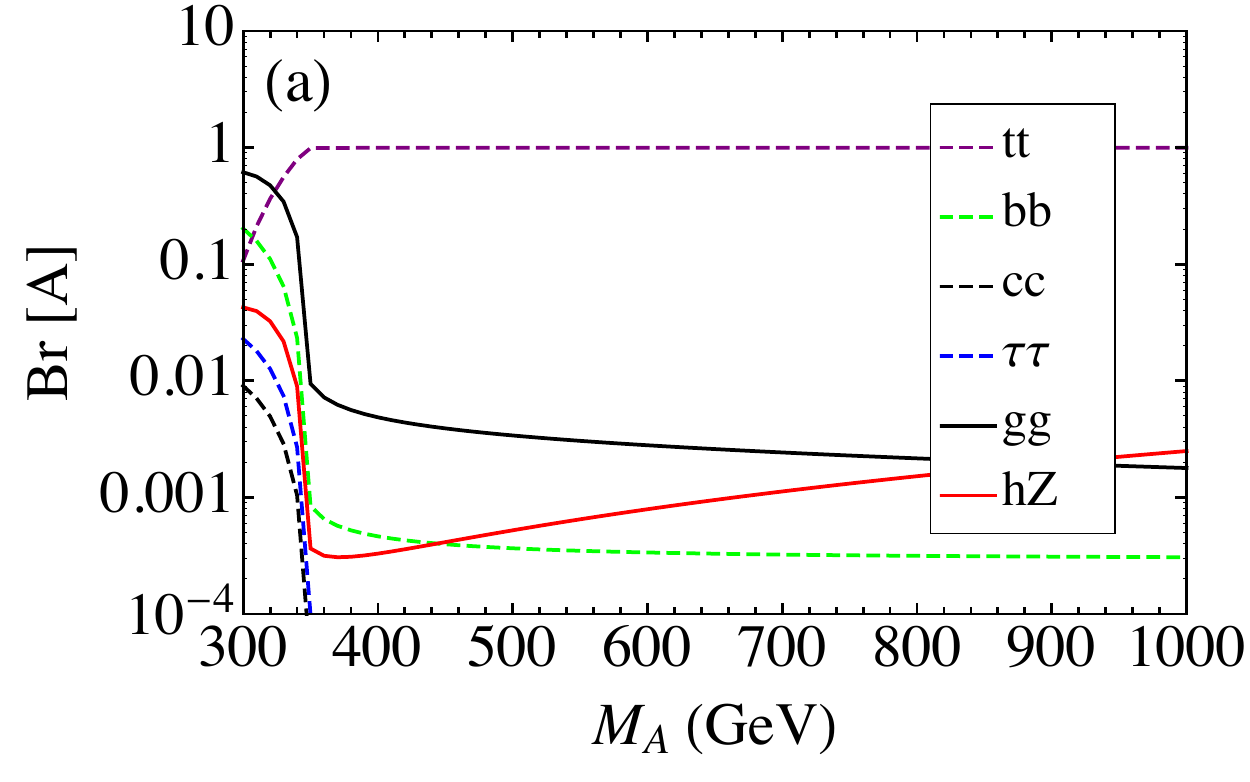}
\includegraphics[width=5cm,height=4cm]{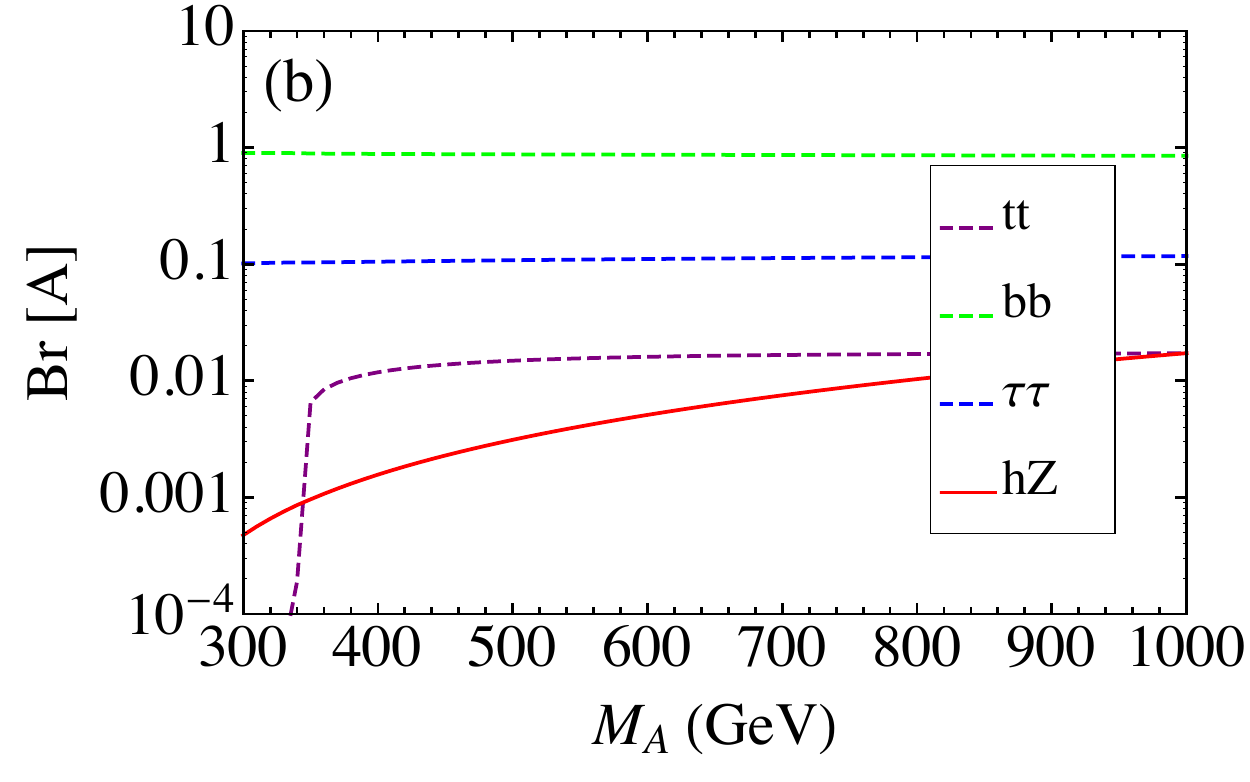}
\caption{\label{fig:BRA} 
The decay branching ratios of the $CP$-odd Higgs boson $A$ for the 2HDM-I with $t_\beta=1$ (upper left), 2HDM-I with $t_\beta = 10$ (upper right), 2HDM-II with $t_\beta =1$ (lower left) and 2HDM-II with $t_\beta =20$ (lower right).
}
\end{figure}

When we relax the alignment limit with the $c_{\beta-\alpha}$ inputs subject to the global fit constraints, the possible decay modes of $A$ in our discussions include $A\to (\bar f f\,,gg\,,hZ)$.
Below, we take the alignment parameters of 
\beqn\label{eq:AhZ_parameter}
&&\textrm{2HDM-I}: c_{\beta-\alpha}=0.2\,,\qquad \textrm{ 2HDM-II}: c_{\beta-\alpha}=-0.02\,,
\eeqn
for the analysis of the $A\to hZ$ decay mode.
Other loop-induced decay widths of $\Gamma[A\to \gamma\gamma]$ and $\Gamma[A\to Z\gamma]$ are typically negligible.
In Fig.~\ref{fig:BRA}, we display the decay branching ratios of the $CP$-odd Higgs boson $A$ in the mass range of $M_A\in (300\,\GeV, 1\,\TeV)$ for the 2HDM-I and 2HDM-II cases, respectively.

\begin{figure}
\centering
\includegraphics[width=5cm,height=4cm]{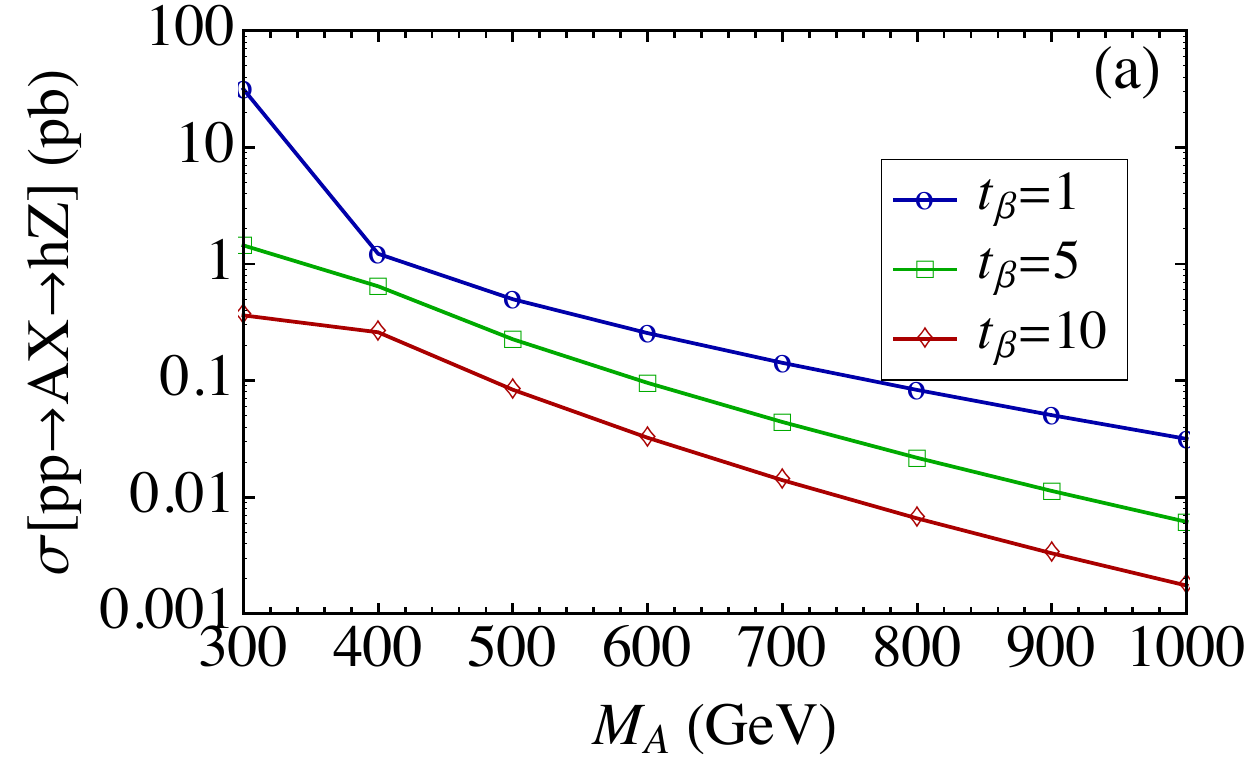}
\includegraphics[width=5cm,height=4cm]{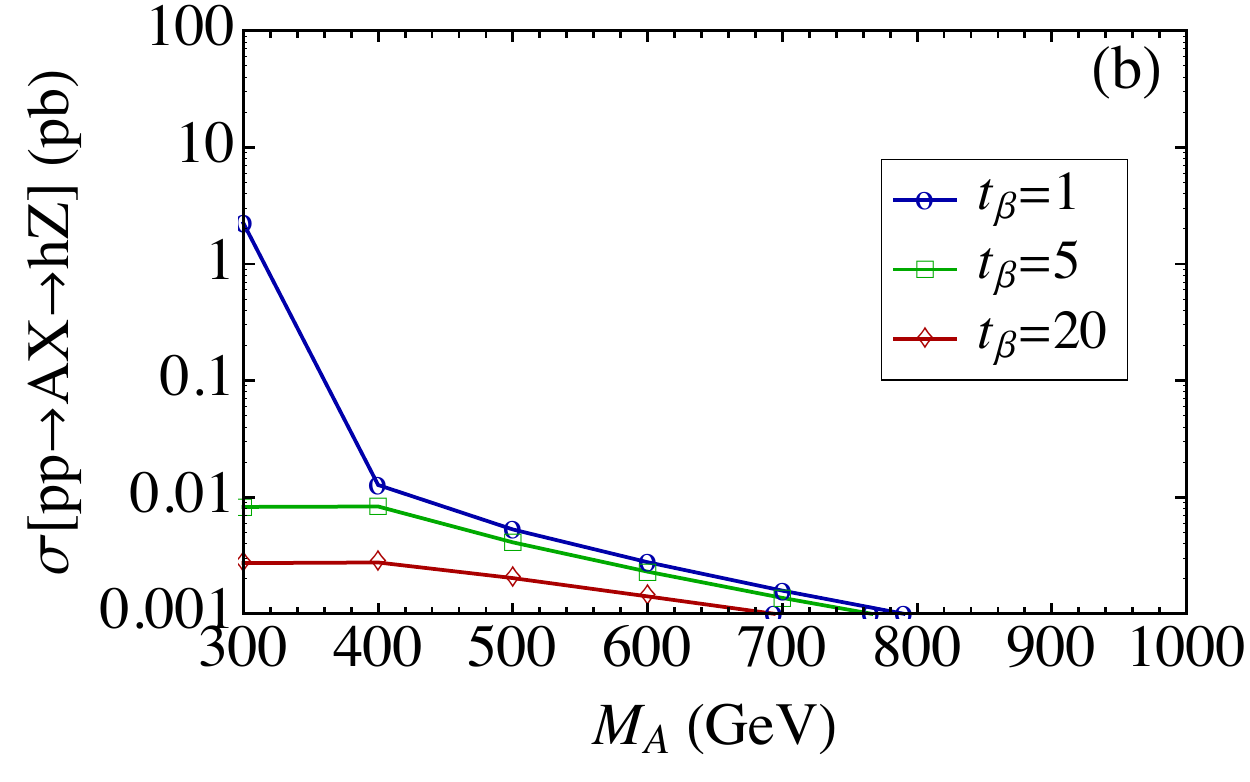}
\caption{\label{fig:xsecA} The $\sigma[pp\to AX]\times {\rm BR}[A\to hZ]$ for $M_A\in (300\,\GeV, 1\,\TeV)$ at the LHC $14\,\TeV$ runs. Left: $M_h=125\,\GeV$ for 2HDM-I. Right: $M_h=125\,\GeV$ for 2HDM-II. }
\end{figure}

Fig.~\ref{fig:xsecA} shows the $\sigma[pp\to AX]\times {\rm BR}[A\to hZ]$ for various cases at the LHC $14\,\TeV$ runs. 
This is done by combining the inclusive production cross sections of $\sigma[pp\to AX]$ displayed in Fig.~\ref{fig:pptoA} and the decay branching ratios of ${\rm BR}[A\to hZ]$ displayed in Fig.~\ref{fig:BRA}. 
In descending order, the curves correspond to input parameters of $t_\beta=1\,,5\,,10$ for the 2HDM-I signal predictions. 
This is largely due to the production cross section dependence on the $t_\beta$ inputs, as shown in Fig.~\ref{fig:pptoA}(a). 
Meanwhile, the corresponding decay branching ratio of ${\rm BR}[A\to hZ]$ for the 2HDM-I case varies moderately in the range of $\mO(0.1)-\mO(1)$, as shown in Fig.~\ref{fig:BRA}. 
Therefore, the search for the $CP$-odd Higgs boson via the $A\to hZ$ channel is possible for the 2HDM-I cases at the LHC $14\,\TeV$ runs, with the integrated luminosities accumulated up to $\mO(10^3)\,\fb^{-1}$. 
In comparison, the signal predictions of $\sigma[pp\to AX]\times {\rm BR}[A\to hZ] $ for the 2HDM-II case are highly suppressed to $\mO(10^{-2})-\mO(10^{-3}) \,\pb$ with $M_A\gtrsim 2 m_t$. 
This is obvious as seen from the more dominant decay modes of $A\to \bar t t$ for the small $t_\beta=1$ input and $A\to (\bar b b\,, \tau^+\tau^-)$ for the large $t_\beta=20$ input, respectively. 
Thus, the search channel of $A\to hZ$ at the LHC $14\,\TeV$ run is of minor interest for the 2HDM-II case.


\section{The Searches for The Heavy Neutral Higgs Bosons via The $t \bar t$ Decay}
\label{section:AHtt}

In this section, we analyze the LHC $14\,\TeV$ searches for the heavy neutral Higgs bosons $A$ and $H$ via the $t \bar t + A/H$ production, with the sequential decay modes of $A/H\to t \bar t $. 
We always tag the top jets $t_h$ by using the {\sc HEPTopTagger} method. 
For the $t \bar t + A/H$ production channel, we shall look for events including a top jet $t_h$ plus SSDL. 
The corresponding SM background processes should include the final states with SSDL plus multiple jets, where a jet may be mis-tagged as the boosted $t_h$. 
Thus, the corresponding SM backgrounds include $t \bar t $, $t \bar t  b \bar b$, $(W^\pm Z\,, ZZ)$ plus jets~\cite{Campbell:2011bn}\,, and $(t \bar t W^\pm\,, t \bar t Z)$~\cite{kang:2014jia}\,. 
%
%
The cross sections of other SM background processes including $(W^\pm W^\pm\,, t \bar t + W^\pm\,, t \bar t + Z)$ plus jets are less than $1\,\pb$.
As we shall show later, the dominant SM background processes after the preselections of $t_h$ plus SSDL are $t \bar t$ and $W^\pm Z$ plus jets. 
Therefore, we neglect all other SM background processes for the $t \bar t + (A/H \to t \bar t)$ signal channel. 
After the reconstruction of the boosted $t_h$, we shall select the kinematic variables for the signal events and carry out the TMVA analysis to optimize the signal significance. 


\subsection{The MC simulations and the top jet tagging}

For event generations of the signal processes, we use Universal FeynRules Output~\cite{Christensen:2008py} simplified models with $A$ or $H$ being the only BSM particles. 
The relevant coupling terms to be implemented are the Yukawa couplings of $A b \bar b/H b \bar b$, and the Yukawa couplings of $A t \bar t /Ht \bar t$. 
We generate events for both signal and SM background processes at the parton level by {\sc MadGraph 5}~\cite{Alwall:2014hca}\,, with the subsequent parton shower and hadronization performed by {\sc Pythia}~\cite{Sjostrand:2006za}\,. 
Afterwards, {\sc Delphes}~\cite{deFavereau:2013fsa} is used for the fast detector simulations. 
In our simulations of both signal and background processes, we include up to two extra jets with the MLM matching in order to avoid the double counting. 
Our fast detector simulations follow the setup of the ATLAS detector.
The Delphes output will be used for the jet substructure analysis by {\sc Fastjet}~\cite{Cacciari:2011ma}\,.

In what follows, we briefly describe the reconstruction of physical objects by the {\sc HEPTopTagger} method. 
The energy flow observables from the Delphes output are used for the jet substructure analysis by {\sc Fastjet}~\cite{Cacciari:2011ma}\,. 
In each event, we cluster the top jets by using the Cambridge-Aachen (CA) algorithm~\cite{Dokshitzer:1997in, Wobisch:1998wt} with certain jet cone size $R_{\rm CA}$. 
By setting the reconstructed top mass range of $ m_t^{\rm rec} \in (140\,\GeV \,, 210\,\GeV)$, the {\sc HEPTopTagger} algorithm finds a candidate boosted top jet which contains three subjets with their total transverse momenta greater than $200\,\GeV$. 
The rate of tagging one $t_h$ can be $\sim 30\,\% - 60\,\%$ with certain choices of $R_{\rm CA}$. 
It is also likely to tag a second boosted $t_h$ at the rate of $\sim 10\,\% - 20\,\%$. For such cases, we always choose the one with the largest $p_T$ as the $t_h$. 
Generally speaking, the tagging rates of top jets vary with different choices of the jet cone sizes $R_{\rm CA}$. 
The boost factors of top jets are enhanced with the heavier resonances of $M_{A/H}$. 
For each signal processes of $pp \to t \bar t + A/H $, we scan the jet cone sizes $R_{\rm CA}\in (1.0\,, 3.0)$ at the step of $0.1$ for reconstructing the top jet $t_h$ in the {\sc HEPTopTagger}. 
In addition, the effects due to the underlying events can be eliminated by the filtering procedure~\cite{Butterworth:2008iy} in the {\sc HEPTopTagger}.
The remaining particles will be clustered into narrow jets by using the anti-$k_t$ algorithm with a jet cone size of $R_{\rm narrow}=0.4$, which are required to satisfy $p_T \ge 20\,\GeV $ and $|\eta|<4.5$.


\subsection{The $t \bar t + A/H$ search results}

For the $t \bar t + (A/H \to t \bar t)$ signal channel, one has four top quarks in the final states. 
After one boosted top quark $t_h$ be reconstructed through its hadronic decay mode, we select events containing SSDL $\ell_1^\pm \ell_2^\pm$ from the semi-leptonic decays of two other top quarks. 
It turns out a significant suppression to the SM background can be achieved by selecting the events containing $t_h$ plus SSDL. 
An example of the preselection efficiencies of events for the $M_{A/H}=500\,\GeV$ case is tabulated in Table.~\ref{tab:MA500eff_14TeV}. 
The suppression rates of SM background events from the $t \bar t$ and $t \bar t b \bar b$ can be as significant as $\sim 10^{-5}$ when imposing the SSDL selection criterion. 
Obviously, the $W^\pm Z$ background becomes the most dominant one after the preselections. 
Meanwhile, one has $\sigma(t \bar t)_{\rm select} \approx 0.1\, \sigma(W^\pm Z)_{\rm select} $ after the preselections. 

 \begin{table}[ph]
\tbl{The preselection efficiencies of the $M_A=500\,\GeV$ (with $R_{\rm CA}=2.1$) signal and background processes at the $14\,\TeV$ LHC. We assume the nominal cross section for the signal process to be $\sigma[pp\to t \bar t + A/H]\times {\rm BR}[A/H\to t \bar t ]=50\,\fb$. 
}
{\begin{tabular}{@{        }ccccccc@{        }} \toprule
  & signal & $t \bar t $ & $t \bar t  b \bar b $ & $W^\pm Z$ & $ZZ$  & $S/\sqrt{B}$    \\ 
 \colrule
 Total cross section $({\rm fb})$ & $50$ & $8.0\times 10^5$ & $3\times 10^4$ & $5.0\times 10^4$ & $1.5\times 10^4$  & $...$ \\
 Preselection of $t_h $ $({\rm fb})$ & $28$ & $3.1\times 10^5$ & $1.4\times 10^4$ & $7.7 \times 10^3$ & $1.9 \times 10^3$  &  $...$ \\
 Preselection of $t_h + $SSDL $({\rm fb})$ & $0.48$ & $0.56$ & $0.11$ & $3.92$ & $0.17$  &  $3.4$ \\ \botrule
\end{tabular} \label{tab:MA500eff_14TeV}}
\end{table}


\begin{figure}
\centering

\includegraphics[width=5cm,height=4cm]{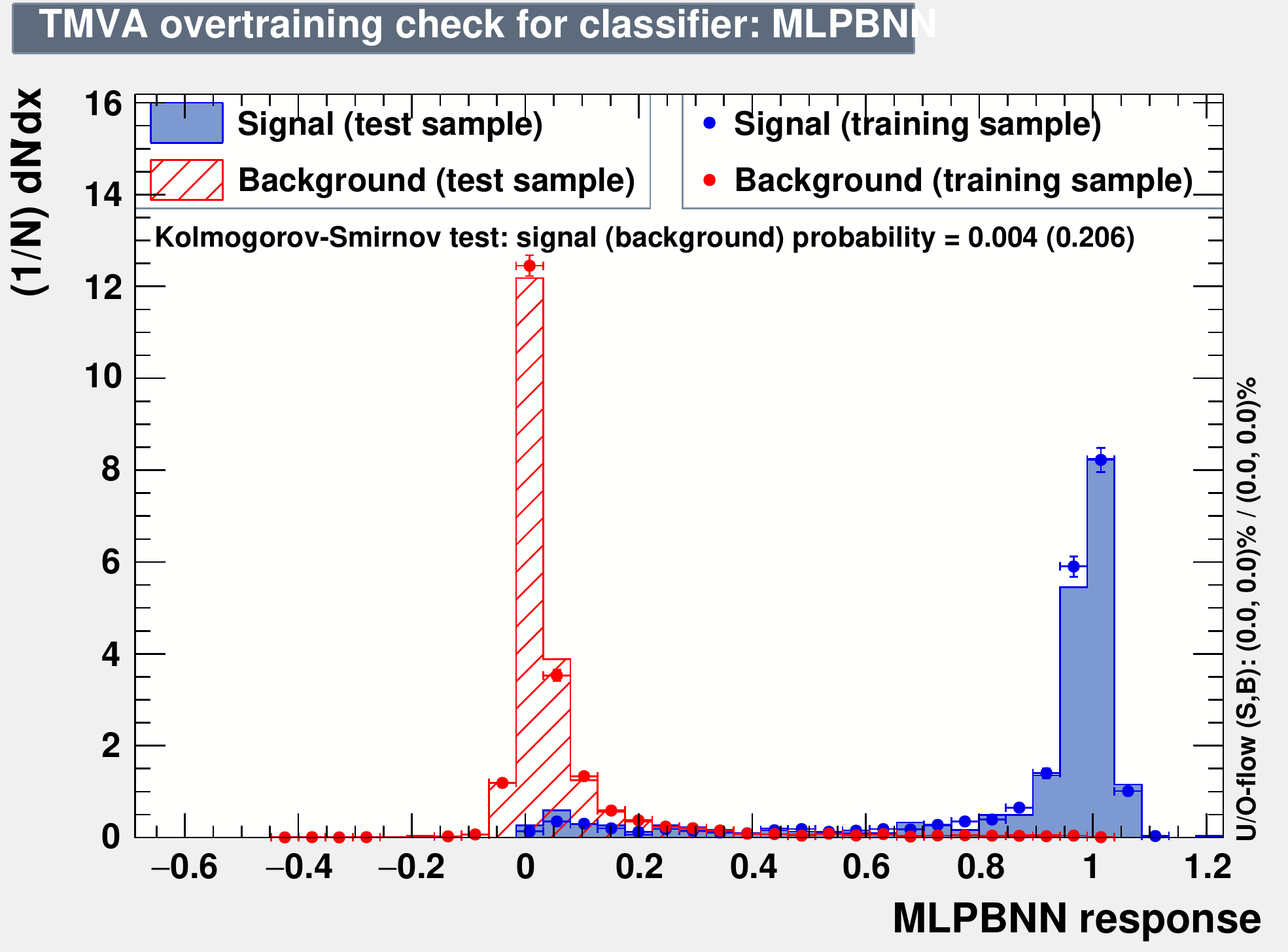}
\includegraphics[width=5cm,height=4cm]{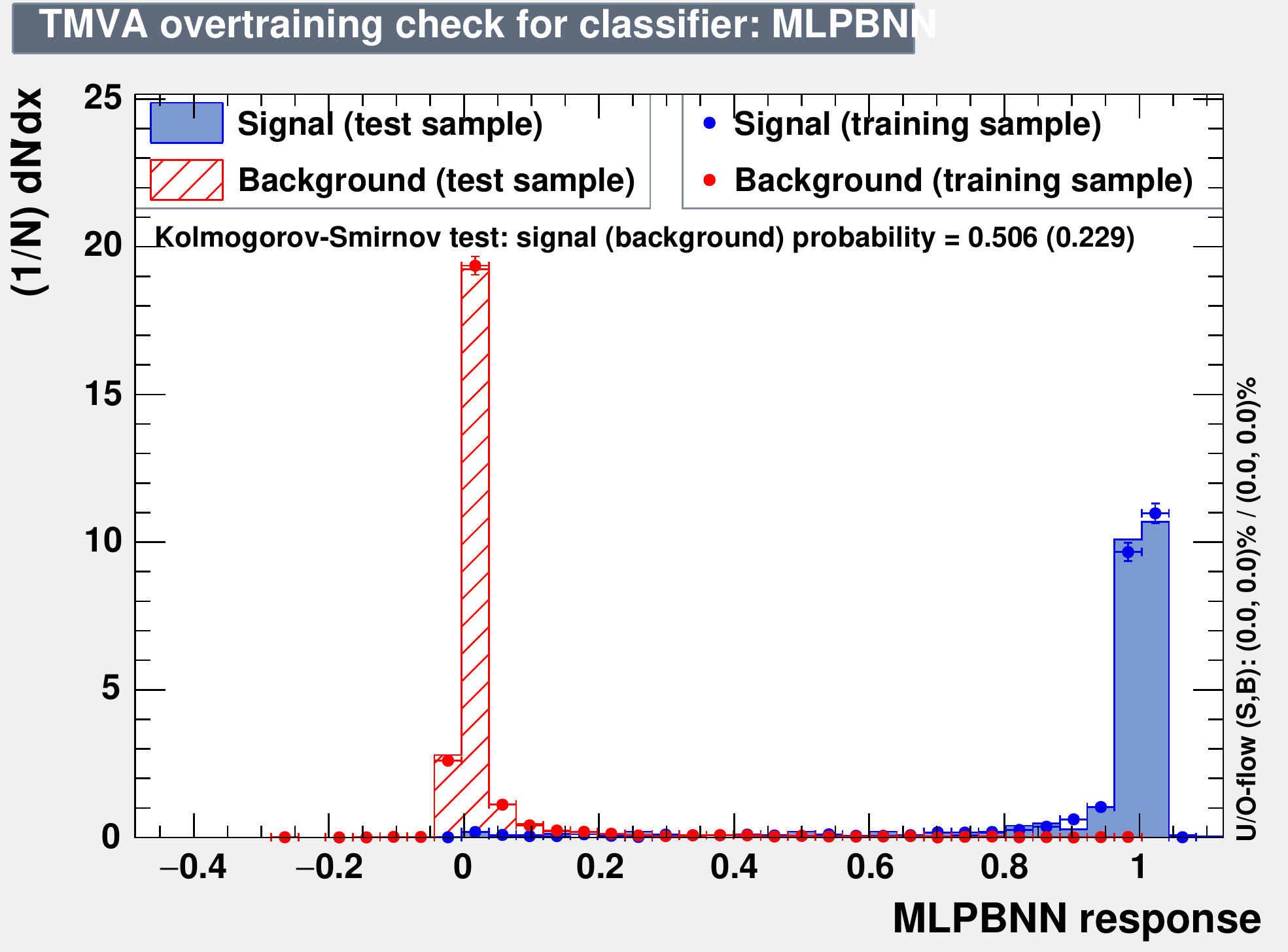}
\caption{\label{fig:ttAH_kin} 
Lower: the normalized distributions of MLP neural network response for signal and background for the $t \bar t + (A/H \to t \bar t )$ channel, left: $M_{A/H}=500 \,\GeV$, right: $M_{A/H}=1000 \,\GeV$.}
\end{figure}

\begin{figure}
\centering
\includegraphics[width=5cm,height=4cm]{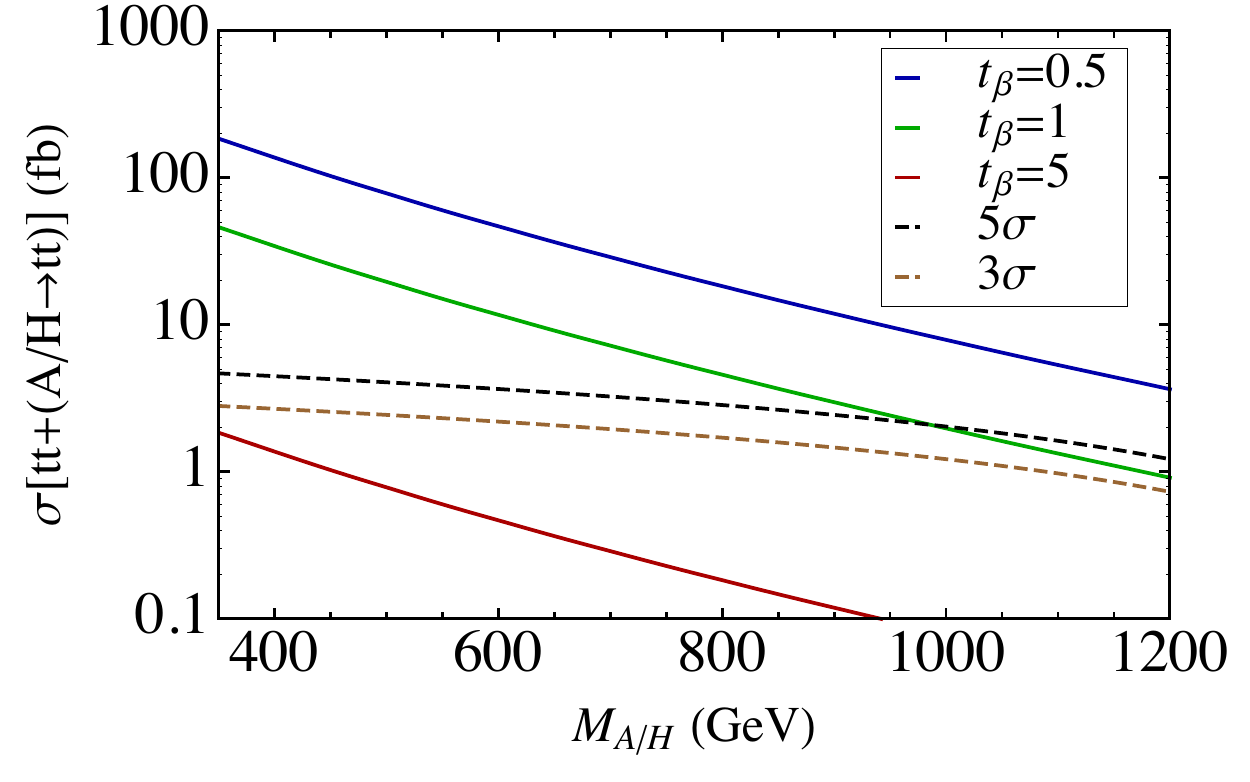}
\includegraphics[width=5cm,height=4cm]{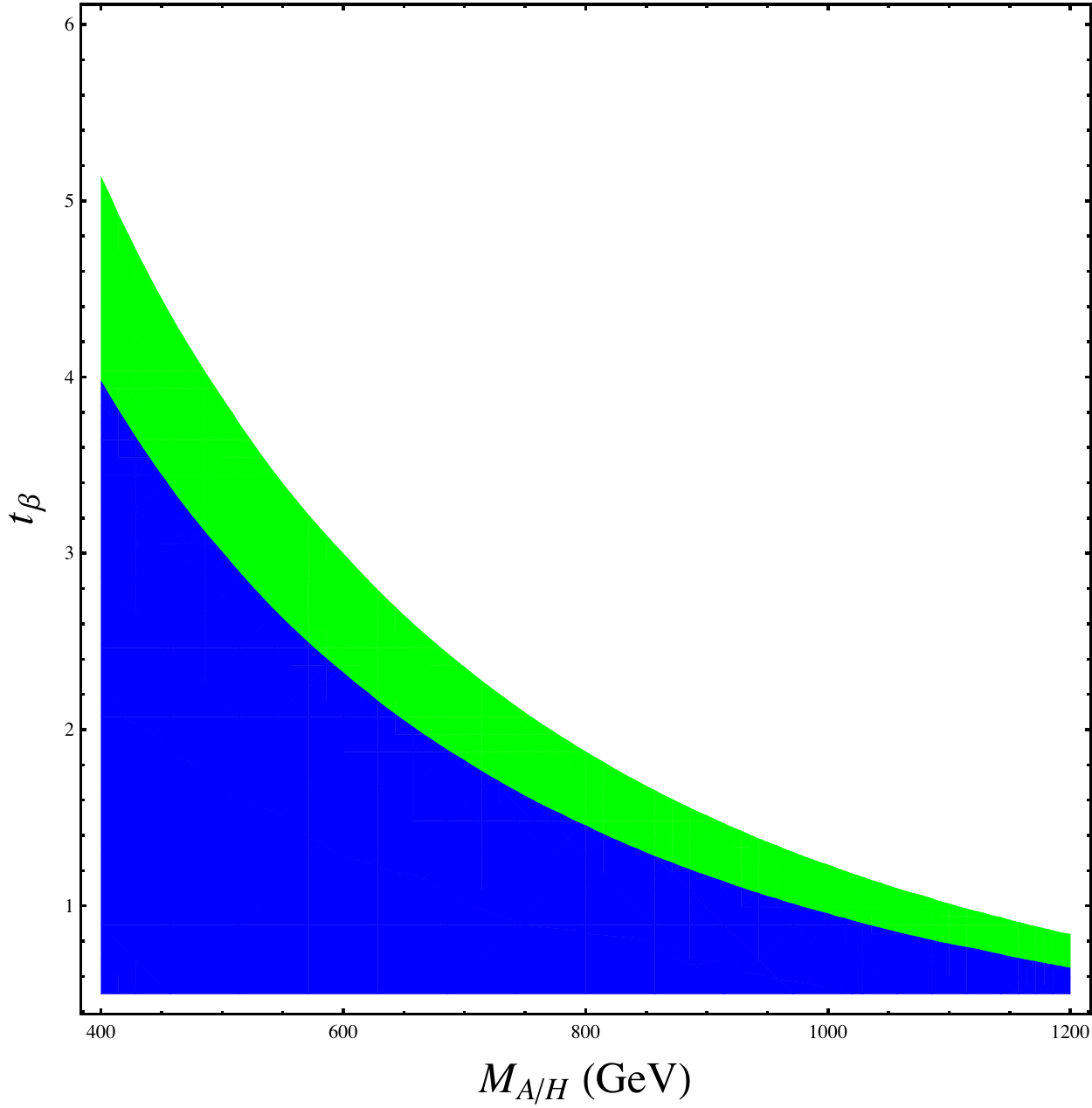}
\caption{\label{fig:ttAHreach} 
The signal predictions together with the signal reaches (dashed lines) of $t \bar t + (A/H \to t \bar t )$ at the HL-LHC runs. 
Left: the mode-independent cross section reaches at the HL LHC runs, right: the $5\,\sigma$ (in blue) and $3\,\sigma$ (in green) signal reaches projected on the $(M_{A/H}\,, t_\beta)$ plane at the HL LHC runs.}
\end{figure}

We perform the multi-variable analysis after the preselections, which is achieved by using the MLP neural network analysis in the ROOT TMVA package.
The list of kinematic variables for the later analysis include: $(p_T\,, \eta\,, \phi)$ of $(t_h\,, \ell_1^\pm\,, \ell_2^\pm)$, $\MET$, number of $( b\textrm{-jets}\,, \textrm{ non-$b$ jets})$, $p_T(b_0\,, j_0)$, $\sum_j p_T(j)$, and $\sum_b p_T(b)$. 
Here $j_0$ and $b_0$ denote the leading non-$b$-jet and the leading $b$-jet ordered in their transverse momenta, respectively. 
The discriminations between signal and background events are shown in Fig.~\ref{fig:ttAH_kin} for the $M_{A/H}=500\,\GeV$ and $M_{A/H}=1000\,\GeV$ samples, which are based on the TMVA output displayed in the Graphical User Interface (GUI).
After obtaining the cut efficiencies, we convert the results to the signal cross sections within the $5\,\sigma$ discovery limits, which read $\sigma[  t \bar t + (A/H \to t \bar t ) ] \sim 2-5\,\fb$ for the mass ranges of $M_{A/H}\in (350\,, 1200)\,\GeV$. 
The results are demonstrated in the left panel of Fig.~\ref{fig:ttAHreach}. 
By looking for the $t_h$ plus the SSDL signals, our analysis shows that the HL LHC searches are likely to reach the heavy neutral Higgs boson masses up to $\mO(1)\,\TeV$ in the low-$t_\beta$ regions for the general $CP$-conserving 2HDM. 
The model-independent signal cross sections for the $5\,\sigma$ reaches are further projected to the $(M_{A/H}\,, t_\beta)$ plane, as shown in the right panel of Fig.~\ref{fig:ttAHreach}.


\section{The Searches for The $A\to hZ$ Decay}
\label{section:AhZ}

In this section, we proceed to analyze the LHC searches for the $CP$-odd Higgs boson $A$ via the decay mode of $A\to hZ$, where we relax the alignment parameter according to Eq.~\eqref{eq:AhZ_parameter}.


\subsection{SM backgrounds and signal benchmark}

The final states to be searched for are the same as the ones in the SM Higgs boson searches via the $hZ$-associated production channel. 
Therefore, the dominant irreducible SM backgrounds relevant to our analysis include $\bar b b \ell^+ \ell^-$, $\bar t t$, $ZZ \to\bar b b \ell^{+}\ell^{-}$, and the $h_{\rm SM} Z\to \bar b b \ell^{+}\ell^{-}$. 
%
%
In our analysis below, we take the $b$-tagging efficiency of $70\,\%$, and the mistagging rates are taken as
\beqn\label{eq:bfake}
&&\epsilon_{c\to b}\approx 0.2 \qquad \epsilon_{j\to b}\approx 0.01\,,
\eeqn
with $j$ representing the light jets that neither originate from a $b$ quark or a $c$ quark~\cite{ATLAS:2012aoa}\,.


\subsection{Jet substructure methods}

Here, we describe the jet substructure analysis and the application to the signals we are interested in. The tracks, neutral hadrons, and photons that enter the jet reconstruction should satisfy $p_T > 0.1\,\GeV$ and $|\eta|<5.0$. The leptons from the events should be isolated, so that they will not be used to cluster the fat jets. The fat jets are reconstructed by using the CA jet algorithm with particular jet cone size $R$ to be specified below and requiring $p_T>30\,\GeV$. Afterwards, we adopt the procedures described in the mass-drop tagger~\cite{Butterworth:2008iy} for the purpose of identifying a boosted Higgs boson:

\begin{itemize}

\item[(i)] Split the fat jet $j$ into two subjets $j_{1\,,2}$ with masses $m_{1\,,2}$, and $m_1 > m_2$.

\item[(ii)] Require a significant mass drop of $m_1 < \mu m_j$ with $\mu = 0.667$, and also a sufficiently symmetric splitting of ${\rm min} ( p_{T\,,1}^2\,, p_{T\,,2}^2) \Delta R_{12}^2 / m_j^2 > y_{\rm cut}$ ($\Delta R_{12}^2$ is the angular distance between $j_1$ and $j_2$ on the $\eta-\phi$ plane) with $y_{\rm cut}=0.09$.

\item[(iii)] If the above criteria are not satisfied, define $j\equiv j_1$ and go back to the first step for decomposition.

\end{itemize}
These steps are followed by the filtering stage using the reclustering radius of $R_{\rm filt} = {\rm min} (0.35\,, R_{12}/2)$ and selecting the three hardest subjects to suppress the pileup effects.

\subsection{Event selection}

The cut flow we impose to the events is the following:

\begin{itemize}

\item[(i)] Cut 1: We select events with the opposite-sign-same-flavor (OSSF) dileptons $(\ell^+ \ell^-)$ in order to reconstruct the final-state $Z$ boson. The OSSF dileptons are required to satisfy the following selection cuts

\beqn
&&|\eta_\ell|<2.5\,,~~~ p_T(\ell_1)\ge 20\,\GeV\,,~~~ p_T(\ell_2)\ge 10\,\GeV\,,
\eeqn
where $\ell_{1\,,2}$ represent two leading leptons ordered by their transverse momenta.

\item[(ii)] Cut 2: The invariant mass of the selected OSSF dileptons should be around the mass window of the $Z$ boson $|m_{\ell\ell} - m_Z|\le 15\,\GeV$.

\item[(iii)] Cut 3: At least one filtered fat jet is required, which should also contain two leading subjets that pass the b tagging and satisfy $p_T> 20\,\GeV$ and $|\eta|<2.5$.

\item[(iv)] Cut 4: Such a filtered fat jet will be then identified as the SM-like Higgs jet. We impose the cuts to the filtered Higgs jets in the mass window of $M_h({\rm tagged})\in (100\,\GeV,150\,\GeV)$.

\item[(v)] Cut 5: We also impose the cuts on the $p_{T\,,h}({\rm tagged})$. 
The SM-like Higgs bosons decaying from the heavier $CP$-odd Higgs boson $A$ would generally be more boosted. 
In practice, we vary the $p_{T\,,h}({\rm tagged})_{\rm cut}\in (50\,\GeV,500\,\GeV)$ and look for the most optimal cuts on $p_{T\,,h}({\rm tagged})$ by counting the corresponding cut efficiencies of $S/B$.

\item[(vi)] Cut 6: Combining the filtered Higgs jets and the tagged OSSF dileptons, the invariant mass of the tagged Higgs boson and the OSSF leptons should reconstruct the mass window of the $CP$-odd Higgs boson $A$: $|M_{h\,,\ell^+\ell^-} - M_A|\leq 100\,\GeV$.

\end{itemize}


\subsection{Implications to the LHC searches for $A$ in the general 2HDM}

\begin{figure}
\centering
\includegraphics[width=8cm,height=4.5cm]{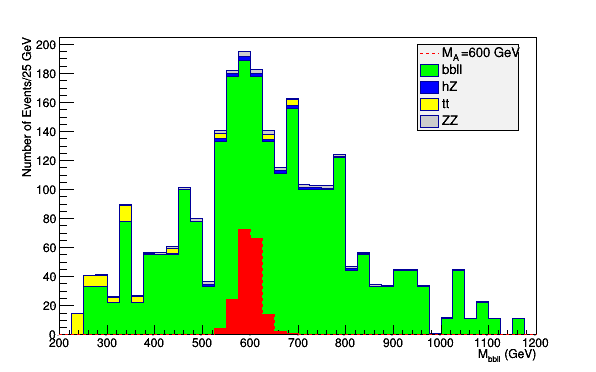}
\caption{\label{fig:AhZ600SB} The $M_{h\,,ll}$ distributions of the $pp \to AX\to hZ$ signal process (for the $M_A=600\,\GeV$ case) and all SM background processes after the kinematic cuts. A nominal cross section of $\sigma[pp\to AX]\times {\rm BR}[A\to hZ]=500\,\fb$ is assumed for the signal. The plot is for the LHC $14\,\TeV$ run with integrated luminosity of $\int\mL dt=100\,\fb^{-1}$. }
\end{figure}

Here, we present the results after the jet substructure analysis and imposing the kinematic cuts stated previously. 
As a specific example of the analysis stated above, the distributions of the $M_{h\,,\ell\ell}$ after Cut 1 through Cut 5 for both signal process and the relevant SM background processes are displayed in Fig.~\ref{fig:AhZ600SB}. 
A nominal production cross section of $\sigma[pp \to AX]\times {\rm BR}[A\to hZ]=500\,\fb$ for the signal process is chosen for the evaluation. 
Among all relevant SM background processes, the $\bar b b \ell^+ \ell^-$ turns out to contribute most after imposing the cuts mentioned above.

\begin{figure}
\centering
\includegraphics[width=5cm,height=3.5cm]{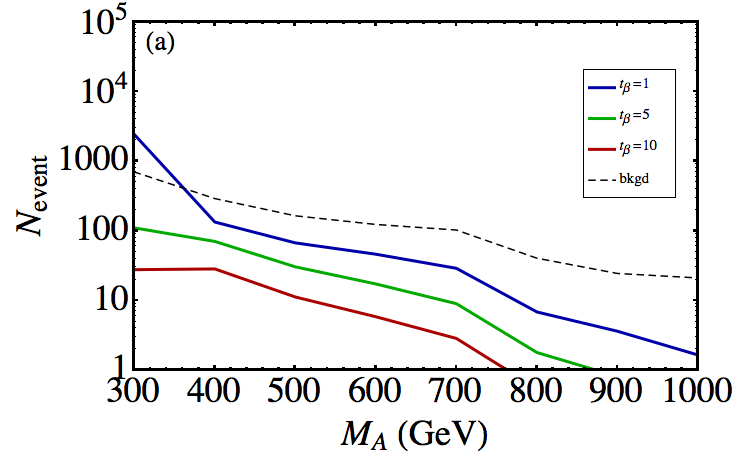}
\includegraphics[width=5cm,height=3.5cm]{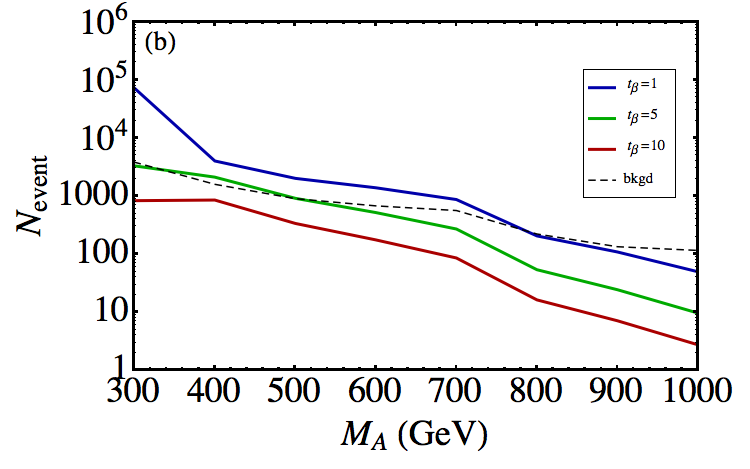}
\caption{\label{fig:AhZevents} 
The number of events for the $pp \to AX\to hZ$ signal in the 2HDM-I after the jet substructure analysis. 
Left: $\int \mL dt = 100\,\fb^{-1}$. Right: $\int \mL dt = 3000\,\fb^{-1}$. 
We show samples with $t_\beta=1$ (blue), $t_\beta=5$ (green), and $t_\beta=10$ (red) for each plot. The discovery limit (dashed black curve) of max$\{ 5\sqrt{B}\,,10  \}$ is demonstrated for each plot.}
\end{figure}

\begin{figure}
\centering
\includegraphics[width=6cm,height=3.5cm]{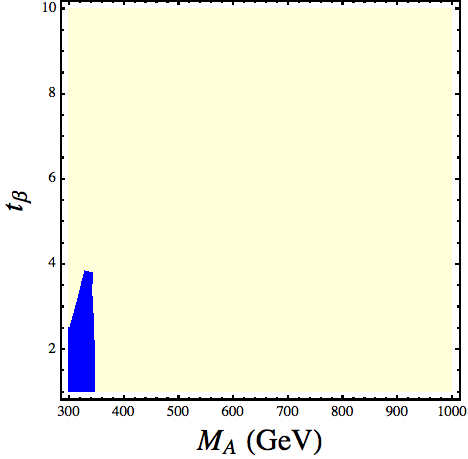}
\includegraphics[width=6cm,height=3.5cm]{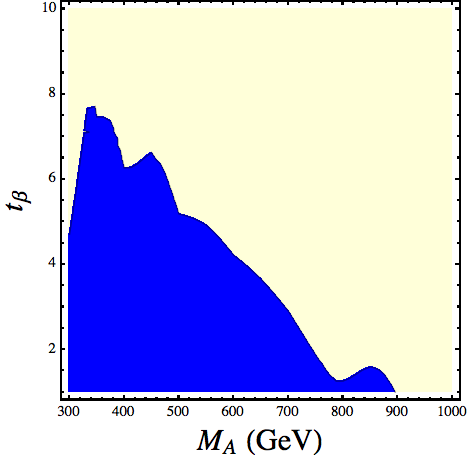}
\caption{\label{fig:AhZIreach} 
The signal reaches for the $A\to hZ$ on the $(M_A\,, t_\beta)$ plane for the 2HDM-I case. 
Left: $\int \mL dt = 100\,\fb^{-1}$, right: $\int \mL dt = 3000\,\fb^{-1}$. 
Parameter regions of $(M_A\,, t_\beta)$ in blue are within the reach for each case.}
\end{figure}

In Fig.~\ref{fig:AhZevents}, we display the number of events predicted by the signal process of $pp\to AX\to hZ$ after the cut flows imposed to the 2HDM-I.
We demonstrate the predictions at the LHC $14\,\TeV$ runs with integrated luminosities of $100\,\fb^{-1}$ and high luminosity (HL) runs up to $3000\,\fb^{-1}$. 
Via the $A\to hZ $ channel, the $CP$-odd Higgs boson with mass up to $\sim 900\,\GeV$ is likely to be probed at the HL-LHC runs.
The signal reaches on the $(M_A\,, t_\beta)$ plane are further displayed in Figs.~\ref{fig:AhZIreach} for the 2HDM-I case.
There are significant improvements of the signal reaches when increasing the integrated luminosity from $100\,\fb^{-1}$ up to the HL LHC runs up to $3000\,\fb^{-1}$. 
For the 2HDM-I case, the $\sigma[pp\to AX]\times {\rm BR}[A\to hZ]$ decreases with the larger $t_\beta$ inputs, as is consistent with what is presented in Figs.~\ref{fig:xsecA}(a) and \ref{fig:xsecA}(b).


\section{Conclusion and Discussion}
\label{section:conclusion}

In this work, we have carried out an analysis of the LHC searches for the heavy neutral Higgs bosons by reconstructing the boosted top quarks and/or SM-like Higgs bosons in their decays.
The decay branching ratios of $A/H \to t \bar t$ can be approaching to $\mO(1)$ with low-$t_\beta$ inputs.
This is the usual case when setting the alignment limit of $c_{\beta-\alpha}=0$ and turning off all possible exotic decay modes of heavy neutral Higgs bosons.
Correspondingly, the searches for the $A/H \to t \bar t$ are of the top priority from the perspective of the production cross sections. 
We consider the $t \bar t + (A/H \to t \bar t)$ signal channel in this work, whose interference effects with the QCD background are less severe compared to the gluon fusion channel. 
In order to suppress the corresponding SM background contributions, we adopt the {\sc HEPTopTagger} method to reconstruct the boosted top jets $t_h$.
As for the $t \bar t  + (A/H \to t \bar t)$ signal channel with multiple top quarks in the final state, we select events containing the boosted $t_h$ plus the SSDL.
Much better signal sensitivity is obtained for this production channel by using the MLP neural network analysis.
For $M_{A/H}\in (350\,\GeV\,, 1200\,\GeV)$, we find that the production cross sections of $t \bar t + (A/H \to t \bar t)$ as small as $\sim [2-5]\,\fb $ can be discovered at $5\,\sigma$ C.L.. 

 With the deviation from the exact alignment limit of the 2HDM, the possible exotic decay mode such as $A\to hZ$ can become significant, especially for the 2HDM-I case. 
This decay channel is due to the derivative coupling term $AhZ$ arising from the 2HDM kinematic terms. 
The technique of the BDRS algorithm of tagging the boosted Higgs jets turns out to be very efficient for suppressing the SM background contributions.
The cut flows to capture the kinematical features for the signal processes were applied thereafter.
The mass reach can be generally up to $\sim 900\,\GeV$ for the 2HDM-I with low-$t_\beta$ inputs at the HL-LHC runs. 
By projecting the sensitivity regions on the $(M_A\,,t_\beta)$ plane for both channel, we find the searches for $t \bar t $ and/or $A\to hZ$ modes can become complementary to the conventional search modes motivated by the MSSM, such as $A/H\to \bar b b$ and $A/H\to \tau^+\tau^+$.


\section*{Acknowledgments}

This work is partially supported by the National Science Foundation of China (under grant No. 11575176), the Fundamental Research Funds for the Central Universities (under grant No. WK2030040069). We would like to thank the conference organizers for the invitation, and the members of the Jockey Club Institute for Advanced Study for their hospitality.


\end{document}